\documentclass[a4paper,twoside]{article}
\usepackage{epsfig}
\usepackage{subfigure}
\usepackage{calc}
\usepackage{lscape}
\usepackage{amssymb}
\usepackage{amstext}
\usepackage{amsmath}
\usepackage{multicol}
\usepackage{pslatex}
\usepackage{fullpage}
\usepackage{psfrag}
\usepackage{comment}
\date{}
\begin{document}
\title{\uppercase{A Secure and Efficient Protocol for Group Key agreement in Heterogeneous Environment}\\
}
\author{{\textmd {\textbf }}
  \fontsize{11}{13}\selectfont
  Mounita Saha, Dipanwita Roy Chowdhury\\
  \fontsize{9}{11}\selectfont
 \textit{Department of compurer science and engineering, Indian Institute of Technology, Kharappur, India}\\
  \fontsize{9}{11}\selectfont
  \textit{mounita@gmail.com, drc@cse.iitkgp.ernet.in}}

\maketitle

\begin{abstract} \fontsize{9}{11}\selectfont Secure group communication in
heterogeneous environment is gaining popularity due to the advent
of wireless and ubiquitous computing. Although a number of
protocols for group key agreement have been proposed, most of them
are not applicable in heterogeneous environment where a number of
computationally limited nodes coexist with one or more
computationally efficient nodes. Among the few existing protocols,
where some fail to satisfy the key agreement properties, some are
unable to handle the agreement for dynamic group. In this work, we
propose a constant round group key agreement protocol for
heterogeneous environment using polynomial interpolation. The
protocol ensures both communication and computation efficiency by
shifting the major computation load on powerful users, achieves
true contributory key agreement property and dynamic handling of
user join and leave. The security of the protocol has been
analyzed under formal model. The comparison result shows
considerable improvement in protocol efficiency compared to the
existing ones.
\end{abstract}



%

%
\section{\uppercase{Introduction}}
\label{sec:introduction}
The key establishment problem has been widely studied in the literature. However, due to the changing scenario of communication applications, it still continues to be an active area of research. The addition of certain protocol properties desired in certain situations and some extra assumptions about the network setup and security infrastructure have opened up new challenges for the key establishment problem. Key establishment is generally classified into two classes: {\it key transport}, where one of the users chooses the key and {\it key agreement}, where all the users contribute to the computation of the key.

In recent times, as different group oriented applications proliferate in modern computing environment, the design of an efficient key agreement protocol for group has received much attention in the literature. One focus area in group key establishment is designing protocols for heterogeneous environment where user nodes with different computation capabilities coexist. Typically in a heterogeneous environment, a number of user nodes have limited computation capability, whereas one or more users have more computation capability. The example of such environment is mobile networks and ubiquitous computing environment.

On the contrary to a common initial impression, secure group communication is not a simple extension
of secure two-party communication. Beyond the fulfillment of security requirements, a large number of the
existing group key agreement protocols suffer from lack of efficiency. Protocol efficiency and scalability in group key establishment is of great concern due to the direct relation of the number of participants to computation and communication complexity.
It can be noted that, one desirable property of GKA in heterogeneous environment is to ensure computation and communication efficiency for the low power users.

In this work, we present a truly contributory group key agreement protocol for heterogeneous environment where a number of resource constrained users are connected to one/more powerful users. Unlike the previous protocols which are based on Diffie-Hellman scheme, our protocol design uses non-Diffie Hellman technique and achieves better computation and communication efficiency. We also present a proof of security of the protocol in random oracle model.

\subsection{Related work}
The original idea of extending the $2$-party key establishment to the multi-party setting
dates back to the classical paper of Ingermarsson et al. \cite{inger82}, and is followed by many works
\cite{tzeng00,steiner00,becker98}.
However, all these approaches simply assume a passive adversary, or only provide an informal/non-standard
security analysis for an active adversary.
Also, in the earlier protocols, the round complexity is linear in the number of group members.

The first constant round protocol secure against passive adversary was given in \cite{bd94}.
More recently, based on this, Katz and Yung \cite{katz03} have proposed the first constant-round protocol for
authenticated group key agreement that has been proven secure against an active adversary. The protocol
requires three rounds of communication and achieves provable security under the Decisional
Diffie-Hellman assumption in the standard model.
While the protocol is very efficient in general, this full symmetry negatively impacts the protocol
performance in a heterogeneous scenario.

In \cite{boyd03} Boyd and Nieto have introduced a one-round group key agreement protocol which is
provably secure in the random oracle model. This protocol is computationally asymmetric.
In recent times Bresson et al. have proposed a number of group key agreement protocols \cite{bresson01,bresson04a,bresson04b} and have given the first provable security model for security analysis of group key agreement protocol.
Bresson and Catalano \cite{bresson04a} have presented a provably-secure protocol
which completes in two rounds of communication. Interestingly, unlike previous approaches, they construct the protocol by combining the properties of the ElGamal encryption scheme with standard secret sharing techniques.
However, this protocol suffers from a significant communication overhead both in terms of the number of messages sent and the number of bits communicated throughout the
protocol. In \cite{bresson04b} another constant round protocol was proposed which is suitable for low power mobile devices. Nam et.al has shown
an attack on it \cite{nam05b}. Then in \cite{nam05a}, Nam et al. proposed a group key agreement protocol for an imbalanced network that provides forward secrecy. In their protocol, the computation time for a mobile node is two modular exponential operations. They adopted the Katz and Yung  scalable compiler to transform their two-round protocol into an authenticated group key agreement protocol with three rounds. However, Tseng \cite{tseng07} later showed that the protocol is not a real group key agreement protocol as the users cannot confirm that their contribution was involved in establishing the group key. \cite{tseng07} also proposed a group key agreement for resource constrained environment which is secure against passive adversary.


\subsection{Our contribution}
The main contribution of this work is to design a contributory group key agreement protocol in heterogeneous communication environment.
Unlike the previous protocols, the proposed protocol at the same time achieves mutual authentication, completes in $2$ round and provides very low computation and communication overhead for the low-power users.
The design goals of a protocol for authentication and key agreement depends on a number of assumptions like the user node capabilities, the communication model setup, i.e. how the users are connected to each other.

\begin{figure}[!h]
\centering
\includegraphics[width=2.2in,height=2in]{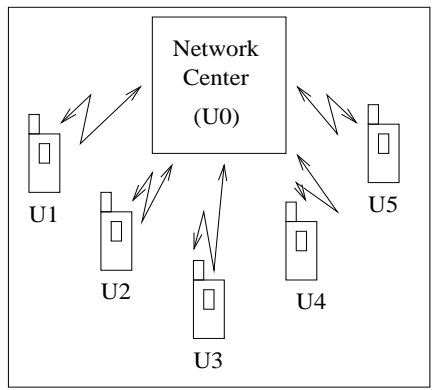}
\caption{System model} \label{model}
\end{figure}

The system model that we consider for this work is shown in figure \ref{model}.
It consists of a cluster of $n$ mobile hosts or users with limited computational
power ${\cal U}=\{U_1,U_2,\ldots,U_n\}$, and a computationally efficient node $U_0$.
The participants communicate with the $U_0$ to establish a common conference key among themselves.
The users do not communicate among themselves.
All the communications are through $U_0$.


The contributions of the work can be summarized as follows:
\begin{enumerate}
\item {\it Asymmetric computation:} In a heterogeneous environment, the computational requirement by the low power nodes can become one major bottleneck if the amount of computation increases with number of users.
In our work, we follow an asymmetric computation pattern and fix the amount of computation required by the host nodes to a constant value. The major computation burden that increases with the number of users are shifted to one/more computationally powerful node.

\item {\it Verifiability of contribution:} In the literature, some protocols \cite{bresson04b,nam05a} have been proposed for server based contributory key agreement both for general and hierarchical layout. However, as pointed out in \cite{tseng07}, none of them assure the user about its participation in key construction and thus user is not able to distinguish between a random key or an actual key.  We note that, the contributory key agreement is meaningful only when the users verify that their contributions are indeed utilized in key construction. In the proposed work, users are able to verify the utilization of their contributions.

\item {\it Efficiency in computation:} We reduce the number of expensive operations required to be performed by each user. Specifically we remove the computationally expensive exponentiation operations and limit the online operations of the users to a single linear function. All other operations are performed offline.

\item {\it Dynamic join and leave :} We consider the users to be completely dynamic i.e. allow the users to leave or join the group within a protocol session.

\item {\it Formal security analysis:} Compared to the number of cryptographic protocols proposed in the literature, security of very few of them have been proved under a formal model. In this work, apart from informal analysis of protocol goals, we provide the security guarantee of the protocols under provable security model.


\end{enumerate}

\section{User-verifiable contributory key agreement}
In this section, we present the proposed group key agreement protocol.

The following notations are used for the protocol descriptions.
    \begin{eqnarray*}
    U &:&\mbox{ The set of users $U_i, i\in(1,n)$}\\
    U_0 &:& \mbox{The leader having higher resources}\\
    ID_i &:&\mbox{ The unique identity of user $U_i \in U$}\\
    {\cal G}_p &:&\mbox{ Cyclic group of order p}\\
    g &:&\mbox{ Generator of group ${\cal G}_p$}\\
    {\cal H} &:& \mbox{ A collision free hash function }\\
    \tau &:& \mbox{A secure signature scheme}\\
    pr_i,pu_i &:&\mbox{Signature key pair for user $U_i$}\\
    C_i&:& \mbox{A counter shared between user $U_i$ and $U_0$ }\\
    \end{eqnarray*}

The public parameters ${\cal G}_p$ and $g$, defined here, are assumed to be known to all the participants in advance. The hash function ${\cal H}$ and the signature scheme $\tau$ is also known to all. Each group member in protocol is having an unique identity $ID_i$. The protocol is defined in an asymmetric setting consisting of a powerful node $U_0$ and a set of group users $(U_1,U_2, \dots, U_n)$. The $U_0$ has a (private,public) key pair $(pr_0,pu_0)$ for encryption-decryption and signature. Each user $U_i$ also has a set of signing and verifying key pair $(pr_i,pu_i)$ for signature generation and verification. Each user $U_i, i\in [1,n]$, shares a counter $C_i$ with $U_0$. The $C_i$ is included for freshness and incremented at each communication session.




\subsection{Proposed protocol}
\begin{itemize}
\item Step 1: {\bf Preparing user contribution and signature}\\
Each user $U_i$ with identity $ID_i$ chooses its contribution $(x_i)$ randomly.
Let $C_i$ be the current value of counter for user $U_i$.
The values of $(ID_i||ID_{0}||x_i||C_i)$ are then encrypted with $U_0$'s public key.
Here $||$ denotes the concatenation operation.

\centerline{$e_i = \{ID_i||ID_{0}||x_i||C_i\}_{pu_{0}}$}

$U_i$ also takes a signature $sig_i$ of $(ID_i||ID_{0}||x_i||C_i)$ using it's private signature key.

\centerline{$sig_i = \tau_{pr_i}(ID_i||ID_{0}||x_i||C_i)$}

Each user then sends $e_i, sig_i$ to the $U_0$.

\centerline{$U_i \rightarrow U_0: e_i,sig_i$}

All these operations can be performed offline. The advantage of using counter over timestamp is that the operations involving the counter can be performed offline.

\item Step 2: {\bf Receipt of user message and verification at $U_0$}\\
The $U_0$ receives all the messages and decrypts them. It
then verifies all the signatures of the  corresponding users. It also
checks the validity of the counter $C_i$ and accepts if the signatures are valid.
\vspace{.1in}

\item Step 3: {\bf Computation of secret by $U_0$}\\
The pair of identity and random value $(ID_i,x_i)$ received from each user is taken as it's contribution to construct the key. $U_0$ also selects a random number $x_0 \in {\cal G}_p$ as its contribution. The secret is constructed by interpolating all the contributions into a polynomial. The $n+1$ values of $(ID_i,x_i)$ are taken as $(n+1)$ input points to the interpolation algorithm. As, all the identities of the users are distinct, a distinct polynomial will be obtained from the fresh input.
Let the coefficients of the resulting polynomial be $a_0,a_1,\dots,a_n$. Thus the polynomial is as follows:

\centerline{$A(x)=a_0 + a_1x + a_2x^2 + \ldots + a_nx^n $}

The secret value is constructed as $K=(a_0||a_1||\ldots ||a_n)$.
\vspace{.1in}

\item Step 4:{\bf Computation of reply message from $U_0$} \\
For each user $U_i$, $U_0$ computes a one way hash ${\cal H}(ID_i,ID_{0},x_i,C_i)$ over the identity $ID_i$, $ID_{0}$, counter $C_i$ and contribution $x_i$.
Then the secret value $K$ is bitwise XORed with this hash value to obtain a value $P_i$ as follows:

\centerline{$P_i= K\oplus {\cal H}(ID_i||ID_{0}||C_i||x_i)$}

If length of $K$ is more than the hash output, it can be sent in multiple fragments.









       Let $Y=\{P_i| i=1... n\}$, $U_0$ takes  a signature $sig_0$ of the values
       $(ID_{0},Y,U)$ using its private signature key.

\centerline{$sig_0 = \tau_{pr_{0}}(ID_{0},Y,U)$}

The $U_0$ finally creates a broadcast message $M=\{Y,U,sig_0\}$ and broadcasts $M$ to all the users.

\item Step 5:{\bf Secret key computation \& Verification at users end} \\
Each user $U_i$ will receive the $U_0$'s messages and  verify the signature of $U_0$.
Then the user obtains the value of ${\cal H}(ID_i,ID_{0},x_i,C_i)$.
This value can be calculated by the user offline.
The shared secret will be calculated by the user as follows:

\begin{tabbing}
\hspace{2in}\=$P_i \oplus {\cal H}(ID_i,ID_{0},x_i,C_i)$\\
\>$=K\oplus {\cal H}(ID_i,ID_{0},x_i,C_i)\oplus {\cal H}(ID_i,ID_{0},x_i,C_i)$\\
\>$=K$\\
\end{tabbing}

If $K$ is sent fragmented, the user has to obtain all the
fragments in a similar manner and combine them to get the secret.


The users can now verify whether the secret is constructed using their contributions. If the contribution $x_i$ of user $U_i$ is used, then the relation $A(ID_i)=x_i$ should be true. The verification is done in the following way:
       After receiving the coefficients user $U_i$ will compute the following

       \centerline{$a_0 + a_1ID_i + a_2ID_i^2 + \ldots + a_nID_i^n$}

       If this value is equal to $x_i$,
       the user knows his/her contribution was  used in key construction. According to Horner's rule, this computation can be written as

      \centerline{$a_0 + a_1ID_i + a_2ID_i^2 + \ldots + a_nID_i^n$}

      \centerline{=$a_0 + ID_i(a_1 + ID_i(a_2 +ID_i( \dots)))$}

       This way, the verification requires only $n$ multiplications.

Finally, the shared secret key for  conference is computed by all the
       users as $Key={\cal F}(K,U)$, where ${\cal F}$ is a predefined one-way function.
\end{itemize}

\begin{figure*}[!t]
\psfrag{e1}{$x_1\in Z_q$}
\psfrag{e2}{$x_2\in Z_q$}
\psfrag{e3}{$x_n\in Z_q$}
\psfrag{e11}{$sig_1=\tau(ID_1,ID_0,x_i,C_i)$}
\psfrag{e22}{$sig_2=\tau(\dots)$}
\psfrag{e33}{$sig_n=\tau(\dots)$}
\psfrag{e4}{$\{ID_1,ID_0,x_1,C_1\}_{pu_s},sig_1$}
\psfrag{e5}{$\{ID_2,.. \}_{pu_s},sig_2$}
\psfrag{e6}{$\{ID_n,.. \}_{pu_s},sig_n$}
\psfrag{e7}{$P_1=K\oplus {\cal H}(ID_1,ID_0,x_1,C_1)$}
\psfrag{e8}{$P_2= \ldots $}
\psfrag{e9}{$P_n=\ldots $}
\psfrag{A}{$A(x)=a_0 + a_1x +a_2x^2 + .. a_nx^n$}
\psfrag{K}{$K={a_0||a_1||\ldots ||a_n}$}
\psfrag{w1}{$P_1=K\oplus {\cal H}(x_i,\dots)$}
\psfrag{w2}{$P_2=K\oplus {\cal H}(x_i)$}
\psfrag{wn}{$P_n=K\oplus {\cal H}(x_i,\dots)$}
\psfrag{K1}{$Y= P1||P2||\dots||Pn$}
\psfrag{Y}{$sig_0=\tau_{pr_0}(ID_0,Y,U)$}
\psfrag{M}{$M=(Y,sig_0,U)$}
\psfrag{e7}{$K=P_1 \oplus {\cal H}(ID_1,ID_0,x_1,C_1)$}
\psfrag{e8}{$K= \ldots $}
\psfrag{e9}{$K= \ldots $}
\psfrag{a1}{$Verify x_1= a_0+a_1ID_1+\ldots $}
\psfrag{a2}{$Verify x_2=.. $}
\psfrag{a3}{$Verify x_n=.. $}
\psfrag{E1}{$Key=F(K,U)$}
\psfrag{E}{$Key=(\ldots) $}
\centering
\includegraphics[width=5in,height=3in]{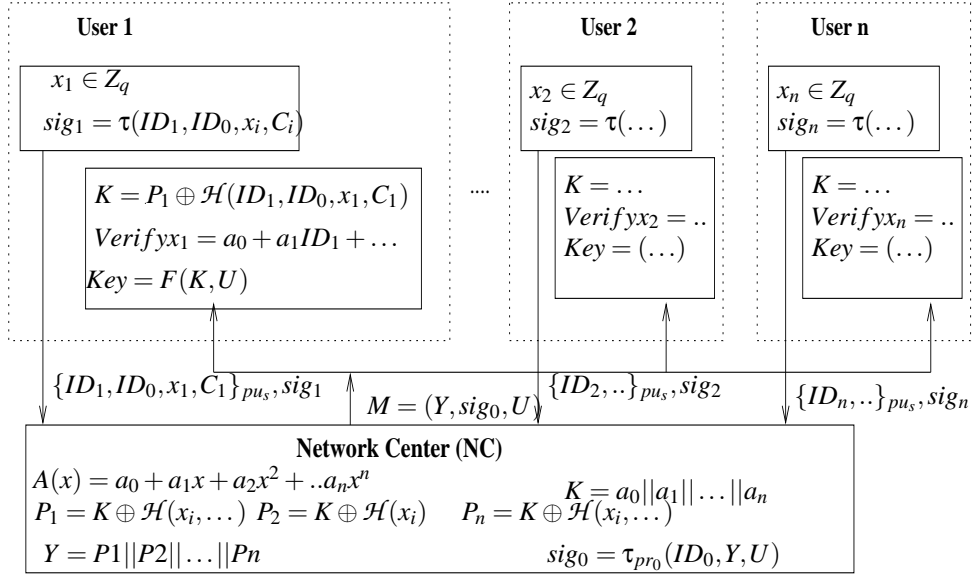}
\caption{Proposed Conference Key Agreement Protocol} \label{poly1}
\end{figure*}

Figure \ref{poly1} demonstrates one instance of the {\it Key\_Agreement} scheme of proposed protocol.

\subsection{Dynamic handling of user join and leave}
When a conference session is in progress, users may be allowed to join or leave. In some applications it may not be desirable that a new joining user understands the content of previous conversations. Similarly, it is also not desirable that a leaving user continues to understand the ongoing conversation. Thus, ensuring the security of the conference while allowing dynamic join and leave is essential. In the proposed protocol, the security of the secret key while maintaining dynamic join/leave is maintained in the following way.

\vspace{.1in}
\noindent
{\bf User Join}\\
When a new user $U_{new}$ joins, it the sends its share $(ID_{new},x_{new})$ to $U_0$. The set of
users is updated as ${\cal U}={\cal U}\cup ID_{new}$. $U_0$ also refreshes its
contribution to $(ID_{0},x'_0)$ using a new random value $x'_0\in {\cal G}_p$. Then the shared secret is computed and distributed as described in steps $3$ to $5$ in key\_agreement.

\vspace{.1in}
\noindent
{\bf User Leave}\\
When an existing user $U_{old}$ leaves, $U_0$ discards its share $(ID_{old},x_{old})$. The set of
users is updated as ${\cal U}={\cal U}\cap ID_{old}$. $U_0$ also refreshes its
contribution to $(ID_{0},x'_0)$ using a new value $x'_0$. Then the shared secret is computed and
distributed as described in steps $3$ to $5$ in key\_agreement.

\noindent
It can be noted that, as one group member joins or leaves, it's corresponding contribution point is added/discarded. The $U_0$'s contribution also changes.
So, whenever there is a change in membership, atleast two points of the secure polynomial change and its value is refreshed. Now, from the property of polynomial interpolation, it is known that, if $1$ out of $(n+1)$ points on a $n$ degree polynomial is changed, the polynomial changes in an unpredictable way. This is information theoretically true.
Thus, secrecy of the previous (new) key from new (former) group members is maintained.

\subsection{Security Analysis}
The proposed protocol has the security properties of key freshness, key confidentiality and mutual authentication. Also true contributiveness of the key is achieved as no participant can predetermine the key or influence the key. An informal analysis shows that it is resistant to common attacks such as replay, impersonation, unknown key share and collusion.

The prime motivation of the proposed protocol is to reduce the computational overhead from the users. Thus, we have deliberately not considered the perfect forward secrecy. However this property can be easily achieved by associating a Diffie-Hellman key exchange.

The advantage of taking identities $ID_i$ as $x$ coordinate values of polynomial interpolation is that they are unique.
However, if the identities of users are known to each other, an user may be able to obtain the contributions of other users. Although this knowledge does not help a new/former user to deduce the old/new key, it may not be desirable in some applications. In that case, instead of using $ID_i$ directly as the $x$ coordinate value, the $H(ID_i,x_i)$ value can be used. As the one way hash is assumed to be collision free, this method will still produce unique values for $x$ coordinates. Alternatively, the counter values $C_i$ known between user and $U_0$ can also be used for $x$ coordinates.

We now present the security analysis of the protocol in formal model.

\noindent{\bf The security model}\\
The first formal model for security analysis of group key agreement
protocols was given by Bresson{\it et al} \cite{bresson04b}. We also use a similar game based security model widely used in literature.


The protocol participants are a set ${\cal U}=(U_0,U_1, \dots, U_n)$ of all users that can participate in the key
agreement protocol. Each user can simultaneously participate in different protocols sessions. Thus an instance of user $U_i$ in protocol session $s$ is represented by the oracle $\Pi_i^s$.
Each user $U_i \in \cal{U}$ obtains a private-public key pair
$(pr_i,pu_i)$ for signature generation/verification.

The {\it partner ID} of an user $U_i$ in session $s$ is
the set of all users who compute the same key as the user $U_i$ in that session. The
{\it partner ID} is defined using {\it session ID}. The {\it session ID} is defined in terms of the messages exchanged among the users in
a session. The detail definition of session identity is given in the \cite{bresson04b}.







\noindent {\bf The adversary}\\
The adversary $\cal{A}$ is active and assumed to have control over
all communication flows in the network.  The
adversary communicates with the users through a number of queries,
each of which represent a capability of the adversary. The queries
are as follows.

\begin{itemize}
\item $Send(U_i,s,m)$: Models the ability of $\cal{A}$ to send message $m$ to user $U_i$.
The adversary gets back from his query, the response that the user
$U_i$ would have generated on processing the message $m$. If the
message $m$ is not in expected format, the oracle would halt. If the
oracle accepts, rejects or simply halts, the reply will indicate
that. If the message $m= NULL$, a new session would be initiated. An
oracle is said to have {\it accepted}, if it has obtained/computed a
session key and accepted it.
\item $Reveal(U_i)$: If an oracle $\Pi_i^s$ accepts and holds a
session key $\cal{K}$, then the adversary $\cal{A}$ can use the
reveal query to obtain the session key held by the oracle.
\item $Corrupt(U_i)$:  When the adversary sends a corrupt query to an user $U_i$,
the internal state information, that the user holds is revealed. Also,
the long term secret key of user $U_i$ is replaced by a value $K$ of
the adversary's choice.
\item $Test(U_i)$: Once an oracle $\Pi^s_i$ has accepted a session key
$K_{ij}$, the adversary can ask a single $Test$ query. In reply to
this query, a random bit $b$ is chosen. If $b=0$ the session key is
returned, otherwise a random string is returned from the same
distribution as the session keys. The advantage of the adversary to
distinguish the session key from the random key is taken as the
basis of determining security of the protocol.
\end{itemize}


\noindent
{\bf Security definitions}\\
Now we define the security assumptions for the proposed key agreement protocol
within the security model given above. The detailed definitions can be found in \cite{bresson04b,nam05a,boyd03}.

\begin{itemize}
\item{\it Freshness}\\
Freshness captures the intuitive fact that
a session key is not obviously known to the adversary. A session key
is fresh if it has been accepted by an uncorrupted oracle and the
oracle or any of its partners are not subjected to the reveal or
corrupt query.


\item {\it Authenticated group key agreement}\\
The security of an
authenticated group key agreement protocol $\cal{P}$ is defined by a
game $G(\cal{A,P})$
 between the computationally bound adversary $\cal{A}$ and protocol $\cal{P}$. The adversary $\cal{A}$ executes the
protocol $\cal{P}$ and executes all the queries described in the security model, as many times as she wishes. $\cal{A}$ wins the
game, if at any time it asks a single $Test$ query to a fresh user
and gets back a $l$-bit string as the response to the query. At a later point of time it outputs a bit $b'$ as a guess for the hidden bit $b$. Let $GG$ (Good Guess) be the
event that $b=b'$, i.e. the adversary $\cal{A}$, correctly guesses the bit $b$.
Then we define
the advantage of $\cal{A}$ in attacking $\cal{P}$, as

\centerline{$Adv_A^P(k)=2.Pr[GG]-1$}

We say that a group key agreement scheme $\cal{P}$ is secure if
$Adv_A^P(k)$ is negligible for any probabilistic polynomial time
adversary $\cal{A}$.



\item {\it Secure Signature Scheme}\\
The security notion for a
signature scheme is that it is computationally infeasible for an
adversary to produce a valid forgery $\sigma$ with respect to any
message $m$ under (adaptive) chosen message attack (CMA).
A signature scheme $\tau(\cal{G,S,V})$ is $(t,q,\epsilon)$ secure if
there is no adversary whose probability in mounting an existential
forgery under CMA within time $t$ after making $q$ queries is
greater than $\epsilon$(negligible). The probability is denoted as
$Succ_{\tau}(\cal{A})$.

\item {\it Secure encryption scheme}\\
 A public-key encryption scheme $PE =(K; E;D)$ consists of three algorithms:
A key generation algorithm $K$ giving a pair $(e;d)$ of matching public and private keys, an encryption algorithm $E$, and a decryption algorithm $D$.




The encryption scheme $PE$ is secure if the adversary's advantage is negligible. We denote the probability as $Succ_{enc}({\cal A})$.
\end{itemize}

Thus, we have defined the security model for the protocol definition. In the next section, we proceed to describe the detail of the proposed protocol.

\vspace{.1in}
\begin{figure*}
\centering
\includegraphics[width=2.7in,height=1.8in]{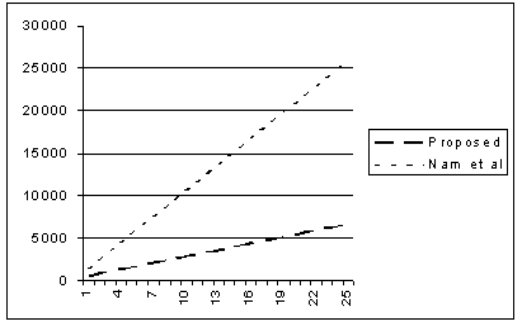}
\caption{Comparison of communication} \label{compare1}
\end{figure*}

\begin{small}
\begin{table*}
  \centering
  \caption{Comparison with existing protocols}\label{compare}
  \begin{tabular}{|c|c|c|c|c|c|c|c|c|}
  \hline
    \centering {\bf Protocol} & User & $U_0$ & round & Message & Dyn & Auth & Verif&PS\\\hline
      \cite{nam05a} & $2$Ex,1sig,1Sv     &$(n+1)$Ex,$1$Sig, $n$ Sv          & $2$ & $n-1$(u),$1$(b) &Y & Y &  N&Y\\ \hline
      \cite{boyd03}& $1$D, $1$Sv &$(n-1)$E,$1$Sig & $1$ & $n$(b)           &N & N&  N&Y\\ \hline
      \cite{bresson04b} &$2$Ex,&$(n-1)$Ex & 2 & $n-1$(u),$1$(b) & Y  &N &N&Y \\ \hline
      \cite{hwang03}& 1Ex,1E&1Ex,$n$D&$4$&$2+3(n-1)$&Y&Y&N&N \\ \hline
      \cite{Jiang06}& 1Ex,1E&1Ex,$n$D&$4$&$2+3(n-1)$&Y&Y&N&N \\ \hline
      Proposed & 1S,1Sv& $(n-1)$D,$1$Sig, $n$ Sv & $2$ & $n-1$(u),$1$(b)&Y &Y &Y&Y\\ \hline
      \multicolumn{9}{c}{Ex:exponentiation Sig: Signature Sv: Signature verification E: encryption D: decryption u:unicast b:broadcast }\\
       \end{tabular}
\end{table*}
\end{small}

\noindent{\bf Proof}\\
We now analyze the security of the protocol as the probability that an adversary can some information on the key and gain some advantage against the authenticated key agreement (AKE) security. Let denote the probability as $Adv_P^{ake}$.
Let ${\cal A}$ be the adversary against the $AKE$ security of the protocol making at most $q_s$ send quires and $q_h$ hash queries (to hash oracles $H$ and $F$). Let ${\cal A}$ plays the game $G_0$ against the protocol.

We now incrementally define a series of games such that each subsequent game has some additional properties. Let $b$ is the bit involved in the $Test$ query and $b'$ be the guess output by the adversary. Then, $Win_i$ denote the event in game $G_i$ when $b=b'$. In each game, we simulate the protocol and consider the adversary to attack the protocol. Finally we relate all of them to obtain the probability of $Win_0$.

Let all the queries are answered by a simulator $X$. It maintains two tables. In the table $S$, it maintains the transcript of all sessions initiated by it. Also, a list $L_H$ is maintained to answer the queries to the hash
oracles. $n$ is the number of users.

\noindent
{\bf Game $G_0$:} This is the real attack. The $X$ generates a pair of signing/verification key and the $U_0$ is given a pair of public-private key. It answers all queries of the adversary in accordance of the protocol.

\noindent
{\bf Game $G_1$:} Let $Forge$ be an event that ${\cal A}$ asks for a send query to the $U_0$ such that the verification of the signature is correct and $m'$ was not previously output by a client as an answer to another send query.  It means that ${\cal A}$ is sending a message that it has produced itself. Such an event can be detected by $X$ as it maintains a table of all protocol transcripts generated by itself. In this case, $X$ aborts the game and outputs $b'$ randomly.

The event $Forge$ occurs when ${\cal A}$ was successful to make an existential forgery against the signature scheme for one of the participants. The probability of this event is thus $n*Succ_{\tau}(\cal{A})$, where $Succ_{\tau}(\cal{A})$ is the success probability of signature forgery against the signature scheme $\tau$, given some public key $PK$.

The game is identical to $G_0$ except when $Forge$ occurs. Thus,

\centerline{$Pr(Win_1) - Pr(Win0) \leq n* Succ_{\tau}(\cal{A})$}

\noindent
{\bf Game $G_2$:} Let $Enc$ be the event when the adversary makes a hash oracle query involving some $(ID_i,x_i)$ and the same hash query was asked by the a protocol participant (user or $U_0$). This can be checked from the list of hash that is maintained. If such an event occurs it means, adversary has been able to attack the encryption scheme.


The probability of success against the encryption scheme after making $q_s$ queries is $q_s*Succ_{enc}({\cal A})$.
The game is identical to $G_1$ except when $enc$ occurs. Thus the total winning probability of the game

\centerline{$Pr(Win_2) - Pr(Win1) \leq q_s*Succ_{enc}({\cal A})$}

Combining all the results, we obtain

\centerline{$Pr(Win_0) \leq  N*Succ_{\tau}({\cal A}) +
q_s*Succ_{enc}({\cal A})$}

Thus, according to our security assumptions, the probability of the polynomially bound adversary to win the game is negligible.

\subsection{Performance analysis}
In this subsection we present a performance comparison of the proposed protocol with the existing ones.

The performance of an authenticated group key agreement protocol is examined based on both its computation and communication requirements. The computation requirement is assessed by the number of major operations performed. The communication requirement is measured by counting the number of rounds, messages and bits to be communicated.

\noindent
{\bf Communication requirement}
In figure \ref{compare1}, we perform a comparison of the communication requirement of the proposed protocol with \cite{nam05a}. The comparison is based on the number of bits required to be transmitted by powerful user $U_0$ versus the number of users. The signatures and hash values are assumed to be $256$ bit whereas the cyclic group of public key system is taken $1024$ bit. It can be noted that the proposed protocol requires much lesser number of bits to be transmitted from the powerful node to the users and the difference grows with increasing number of users. The difference in the proposed protocol is achieved by using non-Diffie-Hellman based key computation technique.

\noindent
{\bf Computation requirement}

Table \ref{compare} shows a comparison of the proposed work with respect to existing similar works. Here first two columns show the computation requirements of user and $U_0$ respectively. Next two columns show the number of rounds and messages required to complete the protocol transactions. The next column (dynamic) denotes whether the protocol is dynamic or not, Auth denotes whether authentication is provided and Verif denotes user verifiability and PS denotes provably secure.



The table shows that the proposed protocol, in-spite of offering the verifiability and mutual authentication property, is comparable to the existing works. Apart from \cite{boyd03}, the rest of the protocols also use $2$ exponentiations. The \cite{boyd03}, despite being computation efficient uses $n$ broadcasts which is expensive. In the proposed protocol, most of the computations performed by the users, i.e. encryption, hash computation and signature can be computed offline. Thus only a bit-wise xor is the main operation to be performed online. Moreover, the offline computations are also less expensive as the user performs a hash computation and one public key encryption which is not expensive as a public key ($3-16$ bit) is short.

\section{Conclusion} In this work we have provided an efficient and scalable solution for true-contributory group key agreement in an heterogeneous environment, which consists of both nodes with limited and relatively higher computational resources. The protocol
transfers most of the computation and communication load to the powerful node, whereas the only online computation performed by a low power user is a single $XOR$ computation.

\end{document}